\documentclass{aa}  

\usepackage{graphicx}
\usepackage{txfonts}
\usepackage[]{hyperref}
\usepackage{xcolor}
\usepackage[normalem]{ulem}

\urldef{\stixflarelist}\url{https://github.com/hayesla/stix_flarelist_science}
\urldef{\stixdatacenter}\url{https://datacenter.stix.i4ds.net/}
\urldef{\ospex}\url{https://hesperia.gsfc.nasa.gov/ssw/packages/spex/doc/ospex_explanation.htm}

\begin{document} 

   \title{Using the STIX background detector as a proxy for GOES}

   \author{Muriel Zo\"{e} Stiefel
          \inst{1}\fnmsep\inst{2}
          \and
          Matej Kuhar \inst{3}
          \and
          Olivier Limousin \inst{4}
          \and
          Ewan C.M. Dickson \inst{5}
          \and
          Anna Volpara \inst{6}
          \and
          Gordon J. Hurford \inst{7}
          \and
          S\"{a}m Krucker\inst{1}\fnmsep\inst{8}
          }

   \institute{University of Applied Sciences and Arts Northwester Switzerland, Bahnhofstrasse 6, 5210 Windisch, Switzerland\\
              \email{muriel.stiefel@fhnw.ch} %\thanks{To be added...}
        \and
            ETH Z\"{u}rich, 
            R\"{a}mistrasse 101, 8092 Z\"{u}rich Switzerland
        \and
            Aircash, Ulica grada Vukovara 271, 10000 Zagreb, Croatia
        \and
            Université Paris-Saclay, Université Paris Cité, CEA, CNRS, AIM, 91191 Gif-sur-Yvette, France
        \and 
            Institute of Physics, University of Graz, 8010 Graz, Austria
        \and
            MIDA, Dipartimento di Matematica, Università di Genova, via Dodecaneso 35, 16156 Genova, Italy
       \and
            Hurford Modulated Solution, 442 Larkin Drive, 94510 Benicia, California, USA
        \and
            Space Sciences Laboratory, University of California, 7 Gauss Way, 94720 Berkeley, USA
            }

   \date{Received October 11, 2024; accepted January 02, 2025}

% \abstract{}{}{}{}{} 
% 5 {} token are mandatory
 
  \abstract
   {The Spectrometer/Telescope for Imaging X-Rays (STIX) onboard Solar Orbiter was designed to observe solar flares in the X-ray range of 4-150 keV, providing spectral, temporal and spatial information. Besides 30 imaging detectors, STIX has two additional detectors, the coarse flare locator (CFL) and the background (BKG) detector, which are used in the present study. Flares observed from Earth are classified using their peak X-ray flux observed by the GOES satellites. Due to the Solar Orbiter mission design, roughly half of all flares observed by STIX are located on the backside of the Sun. These flares lack a GOES-class classification.}
   {In this paper, we describe the calibration of the BKG detector aperture sizes. Using the calibrated measurements of the BKG detector, we explore the relationship between the peak flux for flares jointly observed by STIX and GOES. This allows us to estimate the GOES flare classes of backside flares using STIX measurements.}
   {We looked at the 500 largest flares observed by both STIX and GOES in the time range February 2021 to April 2023. Aperture size calibration is done by comparing 4-10 keV counts of the BKG detector with the CFL measurements. In a second step, we correlate the calibrated STIX BKG peak flux with the GOES peak flux for individual flares.}
   {We calibrated the BKG detector aperture sizes of STIX which are now ready to be implemented into the ground-software (GSW) of STIX. Further, we showed that for the larger flares (C class and above) a close power law fit exists between the STIX BKG and GOES peak flux with a Pearson correlation coefficient of 0.97. This correlation provides a GOES proxy with a one sigma uncertainty of $\approx$ 11\%. We were able to show that the BKG detector can reliably measure a broad range of GOES flare classes from roughly B5 up to at least X85 (assuming a radial distance of 1AU), making it an interesting detector-concept for future space weather missions. The largest flare observed by STIX to date is an estimated X16.5 $\pm$ 1.8 backside flare on the 20 Mai 2024.}
   {}


   \maketitle
%
%-------------------------------------------------------------------

\section{Introduction}\label{Intro}

    The Spectrometer/Telescope for Imaging X-rays \citep[STIX;][]{Krucker_2020} is one of ten instruments onboard the satellite Solar Orbiter \citep{Muller_2020}. STIX observes the Sun in the hard X-ray range covering energies between 4-150 keV. This allows us to study the hottest ($\gtrsim$ 10 MK) plasma and understand the distribution of nonthermal electrons in flares. STIX is designed to measure temporal, spatial and spectral information from flares. For imaging, STIX uses an indirect system based on the measurement of 30 visibilities in the u,v-space by 30 detectors \citep{Massa_2023, Krucker_2020}. These detectors also measure the energy of the incoming photons providing spectral information. Besides the 30 imaging detectors, STIX has two additional detectors, the coarse flare locator (CFL) and the Background (BKG) detector. In contrast to what might be expected from its name, the BKG detector measures more than just the instrument's background flux. It is also able to pick up relevant signal coming from the flare itself. The BKG detector is the only detector on STIX not covered by the attenuator allowing it to monitor the X-ray flux in the lower energies ($\lesssim$ 12 keV) even during the largest flares. All other detectors are covered by the attenuator which gets inserted and removed automatically to control the total count rate. Inserted, the attenuator reduces the flux of the low energy X-rays for the covered detectors \citep{Krucker_2020}. Once calibrated, the BKG detector can therefore be used for spectral and temporal analysis of flares with inserted attenuator. This is of particular interest in our current state in the solar cycle which held several flares with attenuator movement during the last months and more are expected to follow in the coming months as we approach another solar maximum. 

    Solar flares are typically classified based on the flux measured by the Geostationary Operational Environmental Satellite (GOES) located at a distance of 1 AU to the Sun within the 1-8 \r{A} band \citep{Garcia_1994}. Using a logarithmic scale with base ten, the largest flares are X-class flares with a flux above 10$^{-4}$ W m$^{-1}$, in descending order they are followed by M-, C-, B- and A-class flares. To fine tune the classification, numbers between 1-9 are used. For X-class flares this scale has no upper limit, allowing for X10-class flares and beyond. The GOES-class classification was initially designed for space weather applications, but today is commonly used by the solar physics community. The classification allows for an initial characterization of the appearance and strength of a given flare laying the foundation for further applications and analysis. With the mission design of Solar Orbiter \citep{Muller_2020}, roughly half of all flares observed by STIX are not directly visible from Earth. Therefore, these flares lack a GOES-class classification. A statistical comparison between STIX and GOES flux allows us to use STIX flux to determine the GOES-class of these backside flares.
        %----------------------------------------------------------------- 
   \begin{figure}
   \centering
   \includegraphics[trim=0 00 0 00, clip,width=8cm]{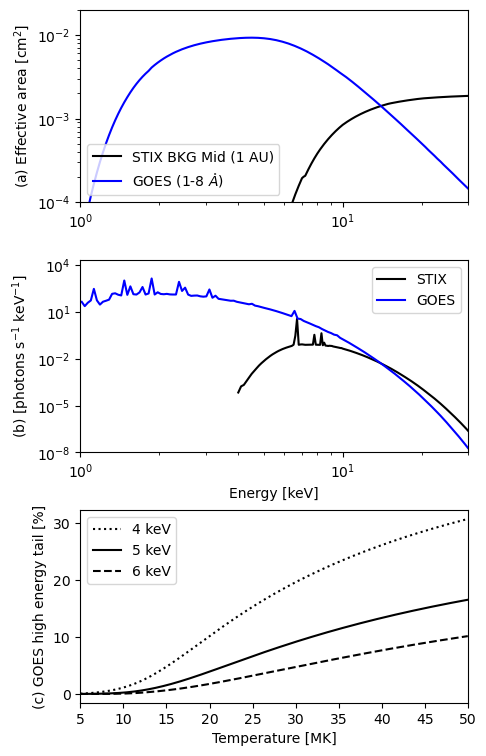}
      \caption{Panel (a) shows the effective area [cm$^2$] as a function of the X-ray energy for the two instruments STIX and GOES. For STIX, the effective area of the two pixels 2\&5 of the BKG detector (see Fig. \ref{Fig: Lightcurve}) is given. Panel (b) shows the hypothetical spectra STIX BKG and GOES would observe for a flare described by an isothermal model with T=20 MK. Panel (c) gives the percentage of high energy photons compared to all photons measured by GOES as a function of the flare temperature [MK]. The three curves show the percentage of the high energy tail of GOES for three different cut-off energies 4 (dotted), 5 (line) and 6 (dashed) keV.}
        \label{Fig: Spectra comparison GOES vs STIX}
   \end{figure}
    %-----------------------------------------------------------------

    Still, when comparing STIX and GOES flux with each other, we need to keep in mind, that the two instruments for the most part do not measure the same flare photons. This is illustrated in Fig. \ref{Fig: Spectra comparison GOES vs STIX}. Panel (a) shows the effective area for the two instruments as a function of the X-ray photon energy. The effective area describes the sensitivity and efficiency of the instrument to detect an incoming photon at a given energy. Note that the effective area displayed here refers only to a part of the STIX BKG detector. The effective area for the whole STIX instrument is roughly 3000 times greater than the curve shown here. In panel (b) we show the hypothetical spectra that GOES and STIX BKG would observe for a given flare with a temperature of T=20 MK. Hypothetical in the sense that GOES does not measure the spectral informations but rather these photons combined make up the actual signal measured by GOES. We used an isothermal model to simulate the photon spectra of a flare [photons s$^{-1}$ keV$^{-1}$ cm$^{-2}$]. Multiplying this by the effective area of the instruments leaves us with the spectra [photons s$^{-1}$ keV$^{-1}$] of the specific detectors. Panel (c) shows the percentage of photons above an energy threshold compared to all photons measured by GOES, we called this the GOES high energy tail, as a function of the flare temperature. The curves of three different energy thresholds, 4, 5 and 6 keV, are displayed. Two observations can be made from Fig. \ref{Fig: Spectra comparison GOES vs STIX}: For one, the high energy tail increases for flares with higher temperatures indicating an increased production by the flare of higher energy photons. Secondly, the percentage is smaller for higher energy cutoffs as GOES is less sensitive to higher energies. STIX only observes photons from 4 keV onward. In an ideal scenario, STIX and GOES would measure the same counts from 4 keV onward. In this case, the black curve shown in panel (c) would represent the percentage of photons measured by STIX compared to all photons measured by GOES. But the transmission factor of STIX in the lower energies ($\lesssim$ 10 keV) is significantly reduced, visible by the drop of the effective area in panel (a). Therefore, the 4 keV cutoff shown in panel (c) represents the upper limit of shared counts between STIX and GOES. However, the actual percentages lie closer towards the curve representing the 5 keV energy threshold. In conclusion, for flares with temperatures in the range 10-20 MK we expect that the STIX photons only correspond to a small fraction of total measured GOES photons, at higher temperatures maybe at best a 10\% overlap is reached. For an additional read on the comparison between STIX and GOES, we refer to the discussion in the Appendix of \citet{Battaglia_2023}.
    
    In \citet{Xiao_2023} a power-law fit between STIX and GOES flux was established using uncalibrated STIX quicklook (QL) data, a low latency data product. As for now, the STIX data-center website\footnote{\stixdatacenter} uses an updated version of this fit. Here, we will use calibrated data of the BKG detector to find the relation between the BKG detector flux of STIX and GOES. This allows us to extend the fit to larger flares, including larger M-class flares and X-class flares for which the attenuator was inserted.

    The paper is structured as follows. In Section \ref{Background Detector chapter} we introduce the BKG detector of STIX in more detail. This is followed by the two main parts of this paper. In Section \ref{Performance of BKG} we present the calibration of the BKG detector apertures of STIX and in Section \ref{GOES and STIX} we compare the flux measured by STIX and by GOES. In Section \ref{Conclusions} we discuss the results and draw conclusions. 

%--------------------------------------------------------------------
\section{The background (BKG) detector} \label{Background Detector chapter}

   \begin{figure*}
   \centering
        \includegraphics[width=18cm]{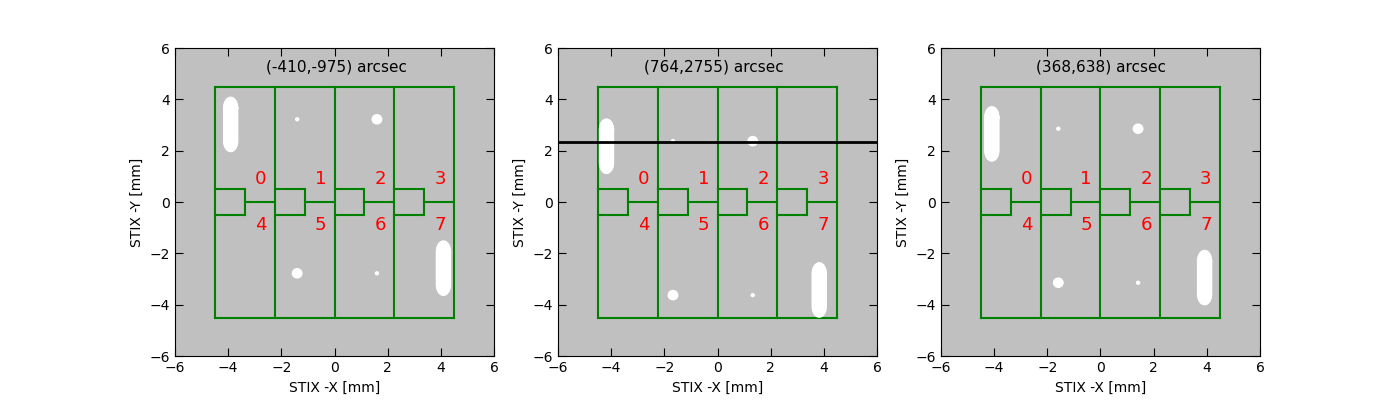}
        \includegraphics[trim=0 0 0 50, clip,width=18cm]{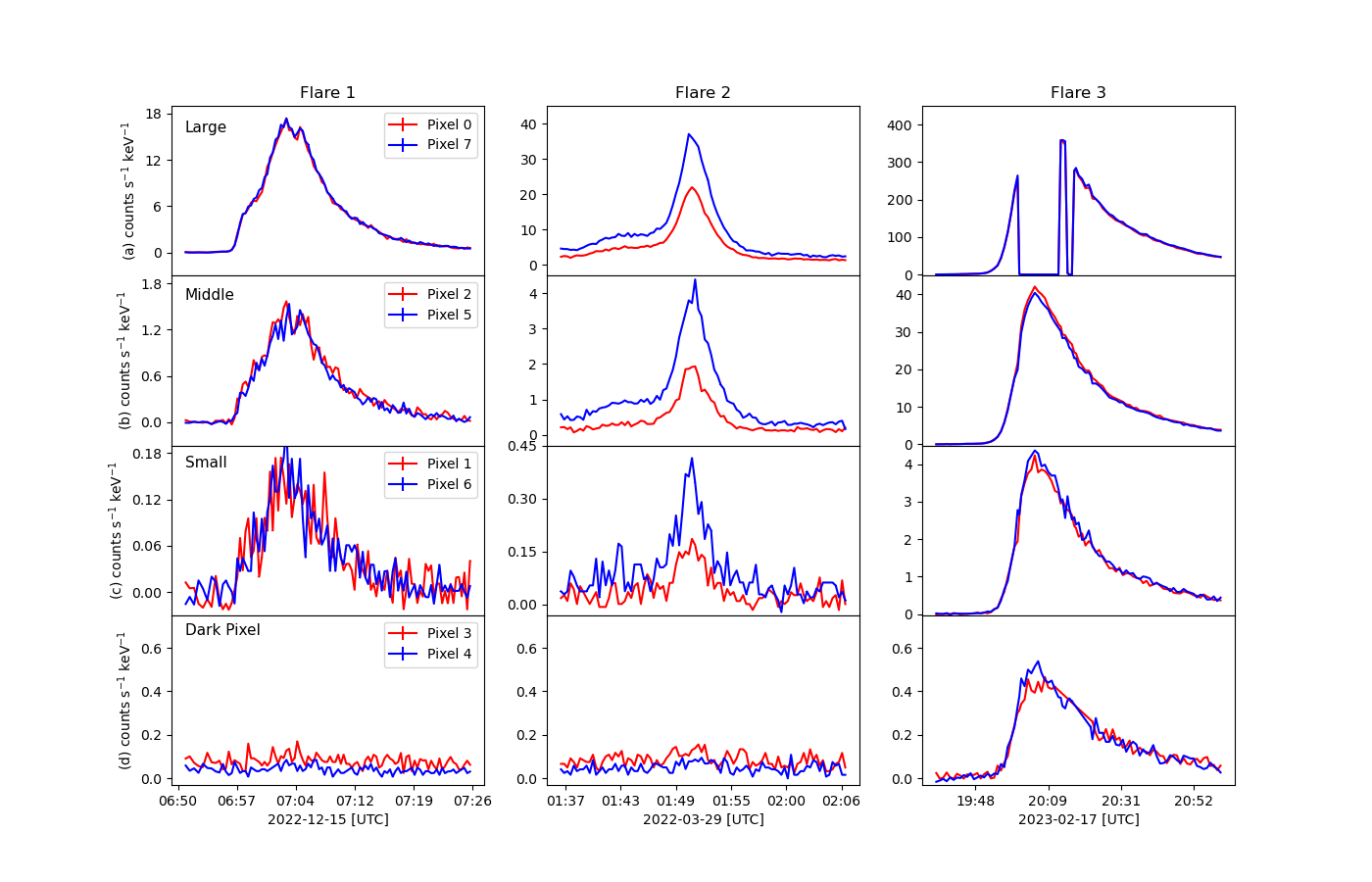}
   \caption{This plot shows the lightcurves of each of the eight pixels (pixel 0-7) from the BKG detector of three different flares in the energy range 4-10 keV. Additionally for each flare, the rear grid of the BKG detector projected onto the pixel structure of the detector is given. The location of each flare in STIX coordinates is given above the respective detector representation. The red numbers indicate the number of the pixel. The black line for flare 2 shows the projection of the boundary of the front grid on to the rear grid. For the lightcurves: Row (a) are the lightcurves of the large apertures (pixel 0 \& 7), followed by the middle (pixel 2 \& 5) apertures in row (b), small (pixel 1 \& 6) apertures in row (c) and last the dark pixels (pixel 3 \& 4) in row (d). From left to right we show three different flares: Flare 1 represents a standard flare, flare 2 is an off-axis flare and flare 3 is a large flare with attenuator movement.}
    \label{Fig: Lightcurve}
    \end{figure*}
%

%-------------------------------------------------------------------

    The BKG detector was designed to serve two main purposes: To measure and monitor the background X-ray flux and to measure the flux of the low energy photons during high flux levels while the attenuator is inserted. The latter is made possible by the fact that the BKG detector is never covered by the attenuator in contrast to the other detectors on STIX. This enables it to monitor the X-ray flux unattenuated during the largest flares. This is especially helpful, when looking at the lightcurve for flares with the attenuator in and for doing spectroscopy analysis of these flares in the low energy range ($\lesssim$ 12 keV). 

    STIX consists of a front grid, a rear grid and the 32 detectors behind the grids. All STIX detectors have 12 pixels, eight large and four small pixels. The large pixels are divided into four top and four bottom pixels. In Figure \ref{Fig: Lightcurve} at the top the green lines outline the 12 pixels as described. For more details on the design, setup and detectors of STIX see \citet{Krucker_2020}, \citet{Limousin_2016} and \citet{Massa_2023}. The grids in front of the BKG detector are arranged as follows: the front grid is entirely open whereas the rear grid is opaque except for six apertures. The nominal sizes of the openings are 1 mm$^2$, 0.1 mm$^2$ and 0.01 mm$^2$ for the large, middle and small ones respectively. The locations and sizes of the apertures in the rear grid for the BKG detector are indicated in Fig. \ref{Fig: Lightcurve} in the top row in white. The relative location between the pixels and the apertures depends on the flare location. The three gray tiles in Fig. \ref{Fig: Lightcurve} at the top show the situation for three different flares. The flare locations given in the STIX reference frame are also displayed. STIX is mounted on Solar Orbiter with a 90 degree rotation. Meaning that the STIX X-coordinate corresponds to positions in north-south directions on the Sun and STIX Y-positions to east-west positions for nominal Solar Orbiter with a roll angle of zero. For each of the three flares in Fig. \ref{Fig: Lightcurve} we plotted the corresponding lightcurves for the eight large pixels sorted by the aperture sizes beginning with the large openings. Flare 1 shows a standard flare for reference. Flare 2 was chosen to serve as a representation for flares which are off-axis. Here, the two pixels with the same opening size do not measure the same flux as one pixel measures significant lower flux. There are two reasons for this discrepancy between the top and the bottom pixels. First, the relative location of the openings in relation to the pixels can move towards the boundary of the pixels. In that case, the pixel does not fully cover the opening anymore, which becomes particularly apparent and important for the large apertures. Secondly, flares which are largely off-axis are not located anymore in the full field-of-view of STIX. In that case, the BKG detector itself is not fully illuminated anymore. We have added a black line in the middle image at the top indicating a projection of the cut of the front grid on to the rear grid. Everything above this line does not receive any direct flux. For the exact definitions of the field-of-views of STIX, see \citet{Massa_2023}, in particular Fig. 12. Flare 3 is a flare with attenuator movement. The attenuator movement is visible in the lightcurves of the large openings. To avoid an excessive increase of the count rate in the detector, these two pixels are automatically switched off and therefore not read out during times when the attenuator is activated \citep{Krucker_2020}.

\section{Calibration of the aperture sizes of the BKG detector} \label{Performance of BKG}

    The detector itself is already calibrated in energy using the calibration sources of STIX. The calibration sources are $^{133}$Ba radioactive source distributed over 128 dots to the sides of the detectors \citet{Krucker_2020}. In this section, we describe the calibration of the aperture sizes of the BKG detector's rear grid. First, we will focus on the measurements which were done on Earth before the mission start of Solar Orbiter which is discussed in Section \ref{Optical estimate}. To confirm these values, we will compare them to the opening sizes estimated from in-flight measurements. For this we will first analyze the ratios between the openings in Section \ref{Analysis 500 flares} before moving on to determine the effective size of the openings in Section \ref{Real Size Estimate}. Finally, we will demonstrate the application of these calibrations in a spectroscopy example. 
    
    %--------------------------------------------------- One column table
   \begin{table}
      \caption[]{Summary of the calibration of the apertures sizes in the BKG detector of STIX.}
      \vspace{-1.8em}
         \label{Tab: Results Calibration BKG Detector}
         $$
         \vspace{-0.8em}
         \begin{array}{lllll}
            \hline\hline
            \noalign{\smallskip}
            \mathrm{Ratio}  & \mathrm{Pixel} & \mathrm{Nominal} & \mathrm{Optical} & \mathrm{In-Flight}  \\
             & & & \\
            \noalign{\smallskip}
            \hline
            \noalign{\smallskip}
            & \mathrm{0/7} & 1.0 & 0.961 \pm 0.007 & 0.9774 \pm 0.0159 \\
            & \mathrm{2/5} & 1.0 & 1.018 \pm 0.043 & 1.0401 \pm 0.0562 \\
            & \mathrm{1/6} & 1.0 & 1.021 \pm 0.156 & 0.9975 \pm 0.1240 \\
            \noalign{\smallskip}
            \hline\hline
            \noalign{\smallskip}
            \mathrm{Size}  & \mathrm{Pixel} & \mathrm{Nominal} & \mathrm{Optical} & \mathrm{In-Flight} \\
            \mathrm{[mm^2]} & & & \\
            \noalign{\smallskip}
            \hline
            \noalign{\smallskip}
             & \mathrm{0} & 1.0 & 1.225 \pm 0.003 & 1.235 \pm 0.146 \\
             & \mathrm{7} & 1.0 & 1.232 \pm 0.003 &  1.307 \pm 0.081  \\
             & \mathrm{2} & 0.1 & 0.111 \pm 0.001 & 0.124 \pm 0.020 \\
             & \mathrm{5} & 0.1 & 0.109 \pm 0.001 & 0.116 \pm 0.019 \\
             & \mathrm{1} & 0.01 & 0.0097 \pm 0.0003 & 0.0106 \pm 0.0019 \\
             & \mathrm{6} & 0.01 & 0.0095 \pm 0.0003 & 0.0112 \pm 0.0024 \\
            \noalign{\smallskip}
            \hline
         \end{array}
         $$ 
         \tablefoot{The top part of the table gives the results for the ratios we got from the analysis described in Section \ref{Analysis 500 flares}. The lower part of the table summarizes the sizes of the apertures as determined in Sections \ref{Optical estimate} and \ref{Real Size Estimate}. For each pixel (or ratio), we report the nominal value, the optical and the in-flight measurement.}
   \end{table}
%

 %-------------------------------------- Two column figure (place early!)
   \begin{figure*}
   \centering
   \includegraphics[width=18cm]{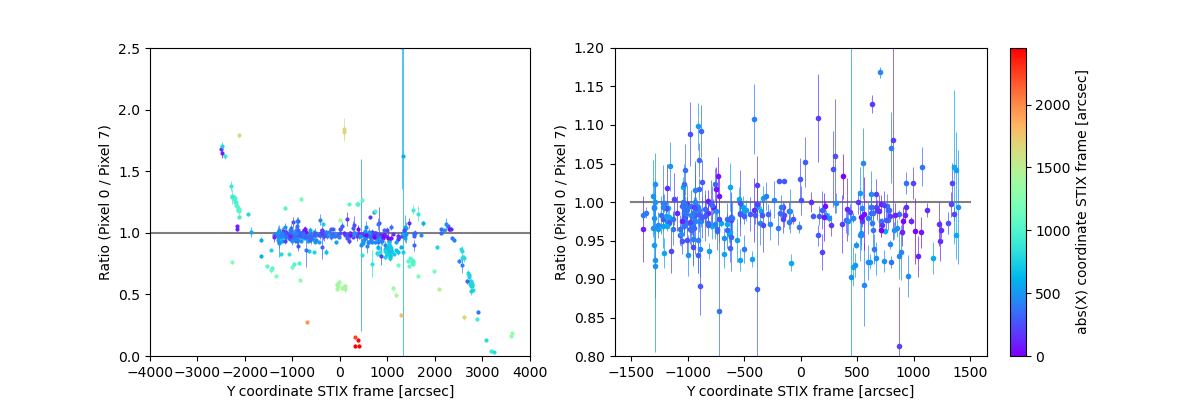}
   \caption{Plot of the ratio between pixel 0 and 7 (pixels behind the largest openings of the BKG detector) as a function of the flare position in the Y-coordinate of the STIX system. The color code corresponds to the flare position in the X-coordinate. The corresponding color bar is given to the right of the image. The left figure shows the result for all 500 flares, the right plot only shows the filtered flares as described in Section \ref{Analysis 500 flares}. The error bars are calculated using error propagation on the uncertainty of the STIX lightcurves. The grey lines show the expected nominal value of the ratio.}
    \label{Fig: Example large apterture}%
    \end{figure*}
%

%--------------------------------------------------------------------

    \subsection{Optical measurement before launch} \label{Optical estimate}
    Prior to the launch of Solar Orbiter, the openings of the rear grid of the BKG detector were measured optically. For this, the instrument Mitutoyo Quickvision 302 at the STIX clean room facility at Paul Scherrer Insitute (PSI) was used. The pixel size of the images is 1.955 microns, the images are stored as 640$\times$480 arrays. This means that the dimension of each image is 1.251$\times$0.938 mm$^2$. While the two smaller apertures (small and middle) fit into one image, three scans have to be stitched together for the largest aperture. The areas were measured by counting the number of bright pixels above a threshold. As the transition from dark to bright in the images taken at 2 micron resolution is rather sharp, the choice of the threshold value is straight-forward \citep{Kuhar_2018}. Error bars of the obtained areas are derived from the spread of plausible threshold values. For the large aperture the areas of the three images were summed up. The results are shown in Table \ref{Tab: Results Calibration BKG Detector} in the fourth row ("Optical").

    \subsection{In-flight analysis of the 500 largest flares} 
    
    \label{Analysis 500 flares}

    To determine the size of the apertures on the rear grid of the BKG detector, we analyzed the 500 largest flares observed by STIX and GOES in the time range of February, 14, 2021 to April 30, 2023. We extracted the flares from the first version of the STIX flare list\footnote{\stixflarelist} by filtering for the largest peak counts in the 4-10 keV QL channel. For the analysis of each individual flare, we used a STIX science and background file. STIX science files contain measurements taken by STIX during a flare while the STIX background files are measurements of the X-ray flux during quiet Sun phases. The latter comprises both the instrument's X-ray background due to the calibration sources and the X-ray background from the Sun. For each flare, the available background file closest in time to that flare is chosen.

    The following analysis was mainly done in IDL using the STIX ground software (GSW) version v0.5.2. For each flare, we extracted the background-subtracted and live-time corrected, 4-10 keV lightcurve of each pixel (pixel 0-7, see Fig. \ref{Fig: Lightcurve}) in the BKG detector individually. STIX detectors operate in a single photon-counting mode. A photon interacting inside a detector triggers the electronics readout process. STIX detectors process only one trigger at a time. During the processing time devoted to reading out the detector, the detector is no longer able to capture a new photon. During this constant dead-time, the arrival of a new photon is ignored and missed. The live-time corresponds to the period of time during which a detector waits for a new event to trigger the detector chain. Live-time correction refers to a correction needed to take into account the statistics of missed photons in order to accurately reconstruct the photon flux, i.e. the flux we would have recorded if the dead-time had been zero. For small flares, the live-time of the STIX detectors remains well above 90\%. The STIX attenuator is inserted when the live-time reaches values around 50\% to restore a live-time close to 100\%. Hence, live-time correction is generally less than a factor of 2. To get the lightcurves we used a modified version of the GSW routine stx\_science\_data\_lightcurve.pro. The routine currently implemented in the GSW, will return the lightcurve for the BKG detector in units of counts s$^{-1}$ keV$^{-1}$ and is explicitly not corrected for cm$^{-2}$. This is highlighted in Fig. \ref{Fig: Lightcurve}, where we see a factor 10 signal difference between the different openings which directly reflects the different opening sizes. Further, we modified the energy look up table (ELUT) corrections compared to the current GSW lightcurve routine which by default does not have any ELUT correction. ELUT corrections adjust the energy channels of STIX compensating for the slight discrepancy between the real measured energy channel of each individual pixel and the predefined STIX energy bins. We implemented the ELUT corrections, ranging from a fraction of one to three percentage, in our modified routine at the edges of the selected energy bin, in our case at 4 and 10 keV. 

%-------------------------------------- Two column figure (place early!)
   \begin{figure*}
   \centering
   \includegraphics[width=18cm]{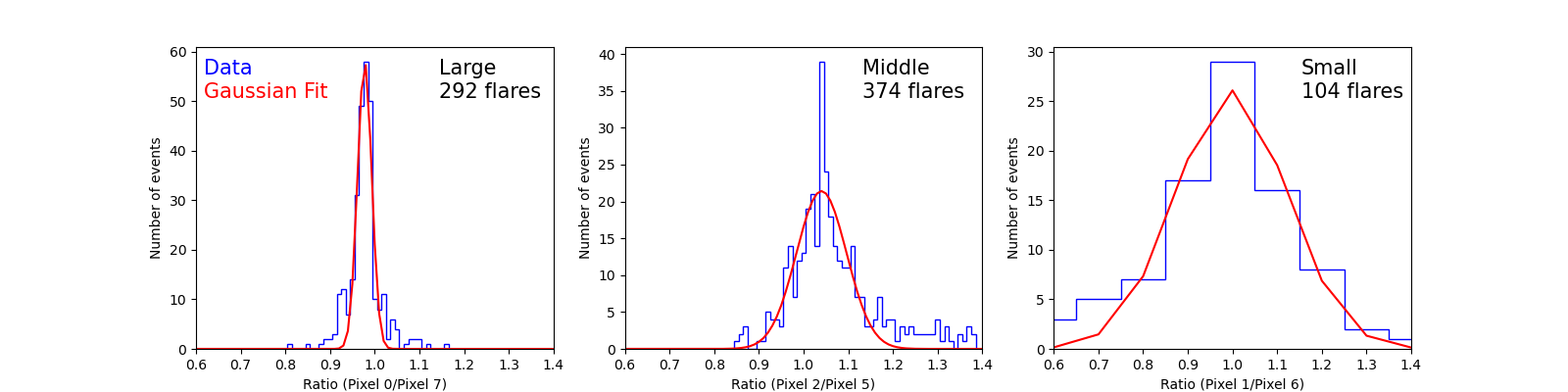}
   \caption{Histogram plots of the ratios between the openings of the BKG detector. From left to right, the plots for the large, middle and small apertures are shown. For each opening, the number of plotted flares is given. The red curve shows the fitted Gaussian curve to the data points.}
    \label{Fig: Histogram all apertures}%
    \end{figure*}
%
%--------------------------------------------------------------------

    From the lightcurves we defined the time range around the flare peak in which the flux is at least 50 \% of the peak flux. This gives us a set of integration times depending on the size and the evolution of the flares. For the majority of the flares, we used flux data from pixel 0 and 7 (i.e. the large apertures) to define the time ranges. However, for roughly 2\% of the 500 flares where the attenuator was inserted, we relied on the data from pixel 2 and 5. For each flare we considered the pixel with the higher peak count to extract the time range.

    Inside these time frames, we determined the mean count $N_i$ for each individual pixel. Between the different pixels we calculated the ratio $R = \frac{N_i}{N_j}$. For example, pixel 0 ($N_i$) was compared to pixel 7 ($N_j$). In Figure \ref{Fig: Example large apterture}, the left panel shows these ratios for each of the 500 flares as a function of the flare location. The turning points where one pixel begins to measure significant more flux than the other are clearly observable. For the large apertures, this indicates an incomplete coverage of the openings by the pixels due to the flare location. Similar turning points are observed for the middle and the small openings as well. However, here this phenomenon is primarily caused by the projection of the front grid onto the rear grid. See also the descriptions in Section \ref{Background Detector chapter} and Fig. \ref{Fig: Lightcurve}. For our further analysis we are focusing on flares that show well-balanced ratios meaning that both pixels receive similar flux while covering the whole opening. Based on the described turning points and the known instrument geometry, we defined the following boundaries for the flare locations: (abs(X), abs(Y)) <  (600, 1400) arcsec for the large apertures, (abs(X), abs(Y)) < (2300, 2000) arcsec for the middle apertures and (abs(X), abs(Y)) < (2300, 2000) arcsec for the small apertures. For the following analysis, we only look at flares within these spatial boundaries. We restricted the flare selection even further for the small and middle apertures, excluding too small flares with insufficient pixel counts where we set the boundaries to the 450 and 130 largest flares for the middle and small openings respectively. Taking both conditions into account, we selected 292, 374 and 104 flares for the large, middle and small apertures respectively for further analysis.

    Figure \ref{Fig: Example large apterture} shows the filtered flares (292) for the large apertures as a function of the flare location. For each of the three different aperture sizes we ensured that the measured ratio of the filtered flares stays constant independent of the flare location to exclude technical artifacts of the production method (stacking of multiple layers). 
    
    The histograms in Fig. \ref{Fig: Histogram all apertures} show the ratios of the different openings. The mean ratio was calculated using a Gaussian fit with the routine "gaussfit" in IDL. This routine applies the following function to fit our data points
    \begin{equation}
        \label{Eq: Gaussian Fit}
        \centering
        f(x) = A_0 \exp^{-z^2/2}
    \end{equation}
    where $z$ is described as $z:=\frac{x-A_1}{A_2}$. The outputs $A_1$ and $A_2$ give us a measure for the mean ratio and its standard deviation. The results are reported in Table \ref{Tab: Results Calibration BKG Detector} under the part "Ratio" in the fifth column ("In-Flight"). For comparison, we calculated the weighted mean and the variance which match the results of our Gaussian fit.

    Moving on, we also looked at the dark pixels, namely pixel 3 \& 4. These two pixels were designed to measure the X-ray background in the instrument itself, created due to the calibration sources of STIX, scattered photons in the instrument and photons produced by solar energetic particle events \citep{Collier_2024}. We made two observations which we add here for completeness. First, the dark pixels measure some signal from the largest flares, as visible in Fig. \ref{Fig: Lightcurve} for flare 3. This is most likely explained by scattered photons reaching the detectors. Therefore, using the number of counts measured on the dark pixels gives a measure for photon scattering during large flares. Looking at our largest flare of the 500, we estimated that the counts measured behind one dark pixel are roughly 0.05\% of the counts compared to the averaged counts of a pixel with a Moiré pattern. Therefore, we do not expect these scattered photons to interfere with imaging or spectroscopy analysis. Secondly, we observed that pixel 3 measures more flux than pixel 4 for the majority of flares. This is e.g. observable in Fig. \ref{Fig: Lightcurve} for flares 1 and 2. A similar effect can be observed on different detectors during non-flaring periods. Our current understanding is that this is likely related to the slightly different individual low energy thresholds for each pixel. However, this effect has no significant impact on imaging and spectral analysis. More details will be discussed in a paper dedicated to the STIX Cadmium Telluride (CdTe) detectors.

    \subsection{Calibration of the BKG apertures}\label{Real Size Estimate}

    %-------------------------------------- Two column figure (place early!)
   \begin{figure*}
   \centering
   \includegraphics[width=18cm]{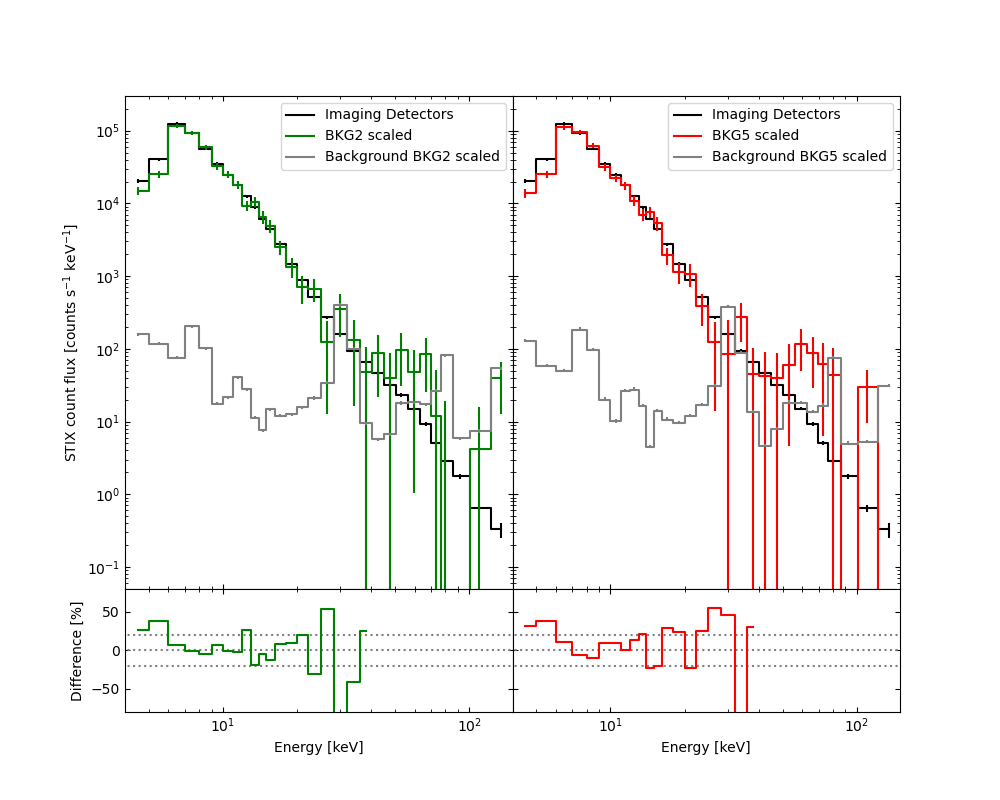}
   \caption{Comparison between the observed spectra of the imaging detectors and the BKG detector of STIX. The spectra is taken during the impulsive phase of the flare shortly before the attenuator moved in. In the left plot we compare pixel 2 and in the right plot pixel 5 of the BKG detector with the 24 imaging detectors behind the coarsest grids. The units of the spectra are counts s$^{-1}$ keV$^{-1}$. To correct for the different areas when using 24 imaging detectors or the individual pixels of the BKG detector, we scaled the BKG detector spectra to the same area as the imaging detectors. In grey the background counts of pixel 2 and 5 are shown in the left and right plot respectively. Below, the relative difference between the imaging detectors and the BKG detector spectra are given.}
    \label{Fig: Spectrosocpy Comparison}%
    \end{figure*}
%
%--------------------------------------------------------------------

    %-------------------------------------- Two column figure (place early!)
   \begin{figure*}
   \centering
   \includegraphics[width=18cm]{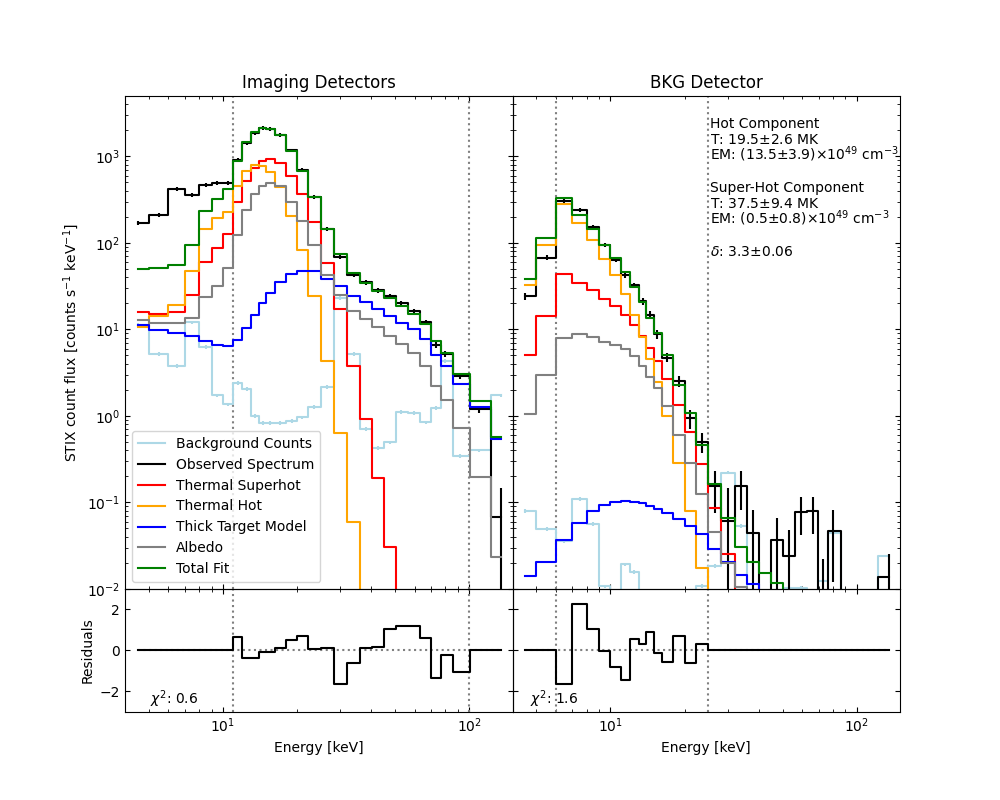}
   \caption{In black, the observed spectra measured by the imaging detectors (left) and the BKG detector with pixels 2 \& 5 (right) are given. The spectra is taken in the impulsive phase during a nonthermal burst with inserted attenuator. For the imaging detectors we fitted the energy range 11-100 keV as denoted by the vertical dotted lines. For the BKG detector we fitted the energy range 6-25 keV. In green the total fit is given with a super-hot (red) and a hot (orange) thermal model, a thick target model (blue) and the albedo (grey). The background counts are given in light-blue. The plots below the spectra show the residuals of the fits.}
    \label{Fig: Spectrosocpy with calibrated BKG detector}%
    \end{figure*}
%
%--------------------------------------------------------------------

    The last step of the in-flight calibration is to determine the size of the openings of the BKG detector's rear grid. Here we use the same method as in Volpara et al. in prep. The special design of the CFL detector (see \citet{Krucker_2020}, Fig. 9 for the exact design) makes it possible for pixels to be fully illuminated for certain flares, meaning that flux reaches the whole area of the pixel. This gives us a measure of the total flux for these flares. We compared and combined the list of flares with fully illuminated CFL pixel with our flare list for further analysis. Using these flares, the following relation gives us a measure of the size of the apertures in the BKG detector:  
    \begin{equation}
        \frac{F_1}{F_2} = \frac{A_1}{A_2}
    \end{equation}
    where $F_i$ is the measured flux and $A_i$ is the illuminated area of the corresponding pixel in the BKG and CFL pixel respectively. For the fully illuminated CFL pixel, we use $A_2=$ 9.282 mm$^2$ as the effective pixel area assuming that 100 \% of this area is illuminated at low energies. This gives us a measure of the area $A_1$ of the BKG detector pixel which corresponds to the size of the aperture.

    As flares with fully illuminated CFL pixels tend to be largely off-axis for STIX, we used the boundaries for the individual openings depending on the flare location as defined in Section \ref{Analysis 500 flares}. This helped us to decide whether we could determine the size of the top and/or the bottom opening. For the two large openings we each had 23 flares with 66 and 65 fully illuminated CFL pixels respectively, for the middle openings each 24 flares with 67 fully illuminated CFL pixels and for the small openings each 7 flares with 19 fully illuminated CFL pixels. In Table \ref{Tab: Results Calibration BKG Detector} we report the mean values we received for the sizes of the individual openings. The error is given by the standard deviation. 

    \subsection{Results calibration BKG detector apertures}

    In Table \ref{Tab: Results Calibration BKG Detector} we report the results from the analysis described in Sections \ref{Optical estimate}-\ref{Real Size Estimate}. Comparing the sizes between the two methods, on Earth optical and in-flight measurements, we see a good agreement for all the openings within the error bars. Using the calibrated sizes for the openings allows us to conduct an exact spectral analysis using the BKG detector.

    \subsection{Demonstration: Application in spectroscopy} \label{Demonstration}

    It was shown in \citet{Caspi_2010} that for large flares, an isothermal model is an unsuitable fit in most cases while a double thermal fit is often more appropriate. Commonly, we differentiate between a hot and a super-hot thermal component of flares. These two components have different physical origins. While the hot component arises from the chromospheric evaporation, the super-hot component most likely stems from the reconnection site in the corona (\citet{Caspi_2010}, \citet{Caspi_2014}). 

    Here we take a closer look at the spectral analysis of the X5-class flare SOL2023-12-31T2140. For more details on the flare in general we refer to \citet{Ryan_2024}. According to \citet{Caspi_2010} we expect to be able to fit two separate thermal components. As the attenuator was inserted during the thermal peak, spectroscopy using the imaging detectors is only possible above 11 keV onward. Using the BKG detector measurements allows us to support the spectroscopy analysis in the lower energies.

    As an initial test, we compared the measured spectrum of the imaging detectors with the spectrum measured by the two middle sized openings of the BKG detector shortly before the attenuator was inserted. Figure \ref{Fig: Spectrosocpy Comparison} shows the comparison between the two. We used the 24 imaging detectors behind the coarsest grids on STIX (see \citep{Massa_2023}). For the comparison between the imaging detectors and the individual pixels 2 \& 5 of the BKG detector, we had to normalize the spectra of the BKG detector to the area of the imaging detector spectra. When comparing the two different spectra they show a high level of congruence in the low energy range except for the 4-6 keV range. This discrepancy in the lowest two energy channels, 4-5 and 5-6 keV, could be due to thickness variations of the 'solar black' coating on the front entrance window of STIX \citep{Krucker_2020}. The transmission of the lower energies strongly depends on the evenness of this layer. Using 24 detectors with 8 pixels each might compensate the irregularities in the thickness of this layer better than two individual pixels in one detector. However, future investigation outside the scope of this paper are need to confirm this. For now we will only consider the spectra of the BKG detector from 6 keV onward for further analysis. Comparing the spectra at higher energies, we see that the BKG detector measures the flare spectra up to 20-25 keV. However, beyond that, the photon counts arriving on the small area of the middle apertures are not sufficient to generate a signal. This indicates that the pixels with openings of the BKG detector measure actual flare spectra at lower energies while detecting mainly background X-ray counts at high energies. The energy limit that separates flare spectrum counts from background counts is variable and depends on the flare size and on the aperture opening size.

    In the next step we derive photon spectra from the observed count spectra. As the STIX spectral response has significant contributions from non-diagonal components (i.e. high energy photons can also make counts at lower energies), the conversion from counts to photons is done by forward fitting. For the following fits, we used the Object Spectral Executive (OSPEX) in the STIX GSW. Within OSPEX, a photon model (e.g. a thermal model) is assumed and convolved with the detector response matrix to compare with the observed count based spectra\footnote{\ospex}. We proceeded as follows to generate a fit combining the 24 imaging detectors and the BKG detector: Initially we used the spectrum from the imaging detectors in the energy range 11-100 keV to get a nonthermal fit. Additionally we estimated an initial double thermal fit. Then we switched to the BKG detector, fitting between 6-25 keV. Using the nonthermal fit from the imaging detectors and fixing it, we fitted the two thermal components. The BKG detector in particular helps to fit the hot component as its spectrum extends down to lower energies. Using this fit, we returned to the imaging detector spectra and overlaid the fit over the spectrum as shown in Fig. \ref{Fig: Spectrosocpy with calibrated BKG detector}. We have found a hot thermal fit with a temperature of T=19.5 $\pm$ 2.6 MK and a super-hot thermal fit with T=37.5 $\pm$ 9.4 MK. These temperatures are in agreement with the range of temperatures seen for large flares as reported in \citet{Caspi_2014}, Fig. 2. Nevertheless, it is important to note that fitting STIX data with the attenuator inserted is often non-unique with different combinations of multi-thermal components giving similarly good fits. Combining the BKG detector and the STIX imaging detectors during attenuated flares should be used to better constrain such a fit. Future efforts should concentrate on jointly fitting spectra from the BKG and the imaging detectors, similarly as has been done with STIX and the Nuclear Spectroscopic Telescope ARray \citep[NuSTAR;][]{Harrison_2013} data \citep{Bajnokov_2024}. 
    
%--------------------------------------------------------------------
\section{Correlating STIX BKG with GOES measurements} \label{GOES and STIX}

    In this Section, we will compare the STIX peak flux measured with the calibrated middle apertures of the BKG detector to the GOES peak flux. This lets us determine the GOES class of backside flares observed by STIX.

    \subsection{Preparation of the STIX data}
    
    For STIX we used the measurements by pixel 2 and 5, which are the pixels covered by the middle sized openings. This makes it possible to compare the flux of the largest flares between STIX and GOES. We excluded off-axis flares for the middle apertures with the boundaries defined in Section \ref{Analysis 500 flares}. For flares that lie beyond these boundaries only the pixel with more counts should be used for flux comparison with GOES flux data. As described in Section \ref{Performance of BKG} we used the live-time and ELUT corrected lightcurves in the energy range 4-10 keV for the 450 largest flares. The choice of this energy range is justified as there is the best overlap between STIX and GOES in this energy range as illustrated in Fig. \ref{Fig: Spectra comparison GOES vs STIX}. For every flare, we extracted the peak value of the lightcurve for both pixel 2 and 5. Then we corrected the flux for the size of the corresponding apertures using the calibrated values (see Table \ref{Tab: Results Calibration BKG Detector}), changing the units of [counts s$^{-1}$ keV$^{-1}$] to [counts s$^{-1}$ keV$^{-1}$ cm$^{-2}$]. Finally, we took the mean of these two values. The STIX flux was then normalized to r'=1 AU using the equation: 
    \begin{equation}
        F' = F\frac{r^2}{r'^2}
    \end{equation}
    where F is the flux at distance r and F' is the flux at distance r'. The distance is given in units of [AU] and the flux is in units of [counts s$^{-1}$ keV$^{-1}$ cm$^{-2}$]. Between April 2023 and April 2024 several flares were observed with attenuator movement which were not yet included into the list of 500 flares used above. To increase the size of our data set of large flares, we included another sixteen flares measured between August 23 to April 24 with STIX attenuator movement and simultaneous observation by GOES. 
    
    \subsection{Preparation of the GOES data}

    For each flare, we downloaded the GOES lightcurve of the low channel 1-8 \r{A} around the peak time of the flare. For each lightcurve we then subtracted the background signal. For this, we used the minimum measured GOES flux 15-60 minutes before the actual flare and subtracted it from the GOES lightcurve. Note that this automatic background subtraction is not well suited for flares occurring immediately back to back. These cases were identified by comparing the GOES flux with the STIX flux, as the GOES flux was disproportionately lower in these cases because to much background flux was subtracted from the GOES lightcurve. For the comparison between STIX and GOES, we extracted the peak value of the GOES lightcurve with units [W m$^{-2}$].

    \subsection{Comparison GOES vs STIX flux}
%-------------------------------------- Two column figure (place early!)
   \begin{figure*}
   \centering
   \includegraphics[width=18cm]{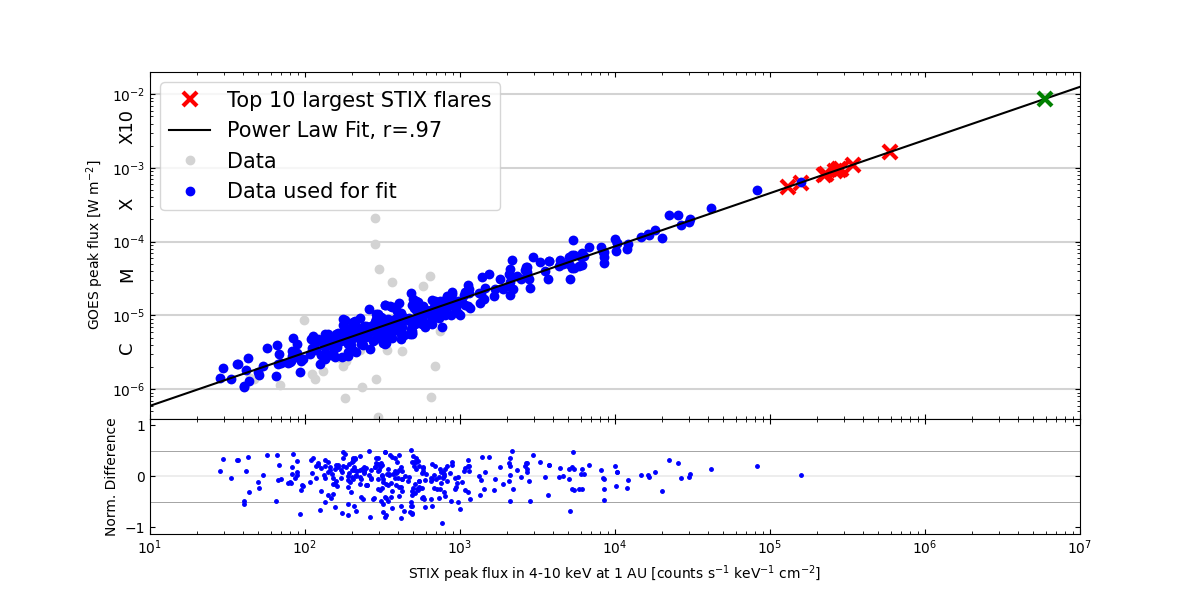}
   \caption{The plots shows the relation between STIX BKG detector peak flux and GOES peak flux. The blue dots represent the flares used for the fit and the black curve is the fit. We have calculated the Pearson correlation coefficient to measure the linearity of the data in logarithmic scale. The grey dots are the outlier flares which were excluded from the fit. The red crosses represent the 10 largest STIX flares in the thermal range up to July 2024. The crosses are plotted using the STIX flux and the fit to estimate the GOES flux. They correspond to the flares listed in Table \ref{Tab: TOP 10}. The green cross marks a conservative upper limit for the largest flare that the STIX BKG detector is able to accurately measure, see the discussion in Section \ref{Conclusions}. In the lower plot we give the normalized difference between the data points (individual flares) and the fit. }
    \label{Fig: GOES vs STIX}
    \end{figure*}
%

%--------------------------------------------------------------------

    Comparing the peak flux of GOES with the peak flux of the BKG detector of STIX, we first focused on the outliers. We were able to cluster the outliers into three groups: flares seen by STIX as a limb flare (i.e. STIX measures disproportionately low flux compared to GOES due to the occultation), flares seen by GOES as a limb flare (i.e. GOES measures disproportionately low flux compared to STIX due to the occultation) or cases where the automatic background subtraction of GOES did not work (i.e. too much background was subtracted for GOES). Excluding these cases from the final STIX and GOES flux comparison leaves us with 354 flares and gives us the relationship shown in Fig. \ref{Fig: GOES vs STIX}. We found the following fit matching the data set: 
    \begin{equation}
    \label{Eq: Result STIX vs GOES}
        y = 10^{0.7213log_{10}(x)-6.948}
    \end{equation}
    where $x=f_{STIX}$ is the measured peak flux of STIX scaled to 1 AU in units [counts s$^{-1}$ keV$^{-1}$ cm$^{-2}$] and $y=f_{GOES}$ is the peak flux of GOES in units [W m$^{-2}$] at 1 AU. To estimate the error of our fit for large flares, we used all flares $\geq$ M1 and calculated the standard deviation of the normalized difference. Smaller flares were excluded for the error estimate, due to their higher sensitivity of the GOES background subtraction. This results in a standard deviation on the GOES flux of 11\% for flares > M1. The individual error bars of each measurement are much lower than the scatter (i.e. the GOES class of a flare is generally much better known than 11\%). This scatter is therefore rather attributed to the differences in the different emission measure distribution (DEM) of individual flares. Depending on the shape of the DEM, the high-temperature bias of STIX is enhanced or suppressed, creating deviations from the fitted curve. As a simple error estimate, it is therefore justified to use the above value of 11\%. For GOES classes significantly larger than the sample of flares used to derive the correlation, the error might increase. We estimated that for flares below X20, the 11\% error clearly dominates. 

    \subsection{Discussion}
    %--------------------------------------------------- One column table
   \begin{table}
      \caption[]{Ten largest flares measured by STIX in the thermal energy range (4-10 keV) up to July 2024.}
      \vspace{-1.5em}
         \label{Tab: TOP 10}
         $$
         \vspace{-0.8em}
         \begin{array}{ccc}
            \hline\hline
            \noalign{\smallskip}
            \mathrm{Date}  & \mathrm{STIX\;flux\;at\;1\;AU} & \mathrm{GOES\;estimate}\\
            \mathrm{[UTC]}& \mathrm{[counts/s/keV/cm^{2}]} & \\
            \noalign{\smallskip}
            \hline
            \noalign{\smallskip}
            \mathrm{2024-05-20} & 594'571 & \mathrm{X}16.5 \pm 1.8 \\
            \mathrm{05:14} &  &  \\
             \hline
             \noalign{\smallskip}
            \mathrm{2024-05-14} & 343'911 & \mathrm{X}11.1 \pm 1.2 \\
            \mathrm{16:47} &  &  \\
             \hline
             \noalign{\smallskip}
            \mathrm{2024-05-15} & 286'764 & \mathrm{X}9.7 \pm 1.1 \\
            \mathrm{08:18} &  &  \\
             \hline
             \noalign{\smallskip}
            \mathrm{2024-07-22} & 280'591 & \mathrm{X}9.6 \pm 1.1 \\
            \mathrm{23:55} &  &  \\
             \hline
             \noalign{\smallskip}
            \mathrm{2024-05-15} & 276'655 & \mathrm{X}9.5 \pm 1.0 \\
            \mathrm{20:37} &  &  \\
             \hline
             \noalign{\smallskip}
            \mathrm{2023-07-17} & 264'387 & \mathrm{X}9.2 \pm 1.0 \\
            \mathrm{00:40} &  &  \\
             \hline
             \noalign{\smallskip}
            \mathrm{2024-07-27} & 230'552 & \mathrm{X}8.3 \pm 0.9 \\
            \mathrm{04:33} &  &  \\
             \hline
             \noalign{\smallskip}
            \mathrm{2024-05-17} & 222'133 & \mathrm{X}8.1 \pm 0.9 \\
            \mathrm{12:27} &  &  \\
             \hline
             \noalign{\smallskip}
            \mathrm{2024-02-22} & 157'433 & \mathrm{X}6.3 \pm 0.7 \\
            \mathrm{22:26} &  &  \\
             \hline
             \noalign{\smallskip}
            \mathrm{2023-07-16} & 129'854 & \mathrm{X}5.5 \pm 0.6 \\
            \mathrm{04:32} &  &  \\
            \hline
         \end{array}
     $$ 
     \tablefoot{The flares are ordered by size. The GOES estimate is calculated using the fit shown in Fig. \ref{Fig: GOES vs STIX}.}
   \end{table}
   
    \citet{Xiao_2023} reported a power law function between the STIX QL lightcurves and GOES. On the STIX data center\footnote{\stixdatacenter} an updated version of this fit is used. This updated version is based on a second order fit and is given by the following equation where x=$log_{10}(f_{STIX})$ and $y=f_{GOES}$
    \begin{equation}
        y = 10^{0.07064x^2+0.07505x-6.9}.
        \label{Eq: Data Center Fit}
    \end{equation}
    Note that in Eq. \ref{Eq: Data Center Fit} $f_{STIX}$ is given in units of [counts s$^{-1}$]. Comparing the fit used by the data center website to our new fit established here, they show a high level of congruence for C and smaller M class flares. For larger flares, however, the GOES estimates on the website start to have large uncertainties as the fit had to be extrapolated to these larger flares. In contrast, the new fit presented here extends the data input all the way to larger X class flares. Another advantage of the fit derived in this paper is the fact that we use science data files instead of uncalibrated QL low latency files. This enables us to do ELUT corrections and detector live-time corrections, which is mainly important for large flares to avoid too low flux. Further we can handle off-axis flares when using science files by choosing the BKG pixel which sees the full flux of the flare. Therefore, we recommend using this improved fit to derive the GOES class for STIX flares, especially for larger flares.

    With the correlation discussed in this paper we can get a GOES proxy from STIX for essentially all flares within the STIX field-of-view ($\sim$2 degrees \citep{Krucker_2020}), as long as we have a flux of at least $\sim$30 counts s$^{-1}$ keV$^{-1}$ cm$^{-2}$ (scaled to 1 AU). For occulted flares the estimated GOES class is lower than the actual GOES class and corresponds to the visible part of the flare as seen from Solar Orbiter.

    The slope of our fit between STIX and GOES is 0.72. This implies that for larger flares, the STIX signal grows more compared to the GOES signal. This aligns with our overall understanding that the flare temperature increase with increasing GOES class (e.g. \citet{Caspi_2014}). To corroborate this statement, we redid the analysis splitting up the energy range 4-10 keV into two intervals. For the energy range 4-7 keV the slope of the fit is 0.75 and for the energy range 7-10 keV the slope is 0.68. For STIX energy ranges at lower energies, the overlap with the GOES energy range increases and the fitted power law index is closer to 1. Contrarily, for the 7-10 keV range, the power law index further decreases as the bias in the sensitivity to high temperatures is further enhanced. As another test, we compared the STIX flux, 4-10 keV, with the long wavelength band 0.5-4 \r{A} of GOES. The slope of the fit between these two data sets is 0.86 confirming our interpretation above. 

    Using the new fit we are now able to determine the GOES class for a wide range of flares observed by STIX, in particular flares observed on the backside of the Sun. In Figure \ref{Fig: GOES vs STIX} the red crosses mark the ten largest flares in the thermal energy range of 4-10 keV scaled to 1 AU measured by STIX up until July 2024. The ten flares are also given in Table \ref{Tab: TOP 10}. The errors on the GOES class estimate are calculated using the 11\% uncertainty on the estimated GOES flux. The only flare recorded by both STIX and GOES on disk is SOL2024-02-22 with an allocated X6.4-GOES class compared to our estimated X6.3-class flare. The good agreement between the two is partially biased through the fact that this flare was part of the data set to determine the fit and has shaped the fit decisive for high X-class flares. The flares on the 14 and 15 of May 2024 were seen by GOES and STIX but all three flares were occulted from Earth's view. The other six flares are backside flares. Three main active regions located on the far side of the Sun in July 2023, May 2024 and July 2024 were responsible for nine of the ten flares in the list. Therefore, the lack of largest flares on the Earth facing side of the Sun is not significant.

\section{Overall discussion and conclusions} \label{Conclusions}
    An in-flight calibration of the BKG detector and its tungsten pinhole collimators enables us to use the BKG detector for spectral analysis in the low energy range ($\lesssim$ 25 keV) even for the largest flares while the attenuator blocks these lower energies from reaching the other STIX detectors. This was demonstrated in Section \ref{Demonstration} using the X5 class flare SOL2023-12-31 as an example. The uncertainty of the opening size measurements is size dependent. We estimated 2 \%, 1 \% and 3 \% uncertainty on the sizes for the large, middle and small apertures.

    We continued by comparing the flux measured by the BKG detector of STIX to the flux measured by GOES to establish a relationship between these two instruments extending the fit to the largest flares. The GOES class is a useful first assessment to characterize a flare. Therefore, the correlation between STIX and GOES is an important tool to estimate the GOES class for flares observed by STIX on the backside of the Sun. We confirmed here a power law function between STIX BKG and GOES. We compared this newly introduced fit with the pre-existing fit using STIX QL-data and confirmed that the fit currently used on the data center provides a good estimate for C and M class flares. However, for larger flares the fit presented in this paper holds several advantages. For one, it is more accurate as lightcurves are used which are not affected by the attenuator. Beyond that, it allows for live-time correction on the measured data. 
    
    As a final point of discussion we want to shortly think about the given name "background detector" for the detector under discussion in this paper. The name "background" implies the measurement of a background flux. As it was discussed in this paper, only the two dark pixels (pixel 3 \& 4) measure always the background X-ray flux in the instrument. While all the large pixels in the BKG detector measure the background flux in the high energy range ($\gtrsim$ 25 keV), the six large pixels with openings in front of them measure additional relevant flare signals in the lower energy range. As demonstrated in this paper, the signal can be used for flare analysis or GOES class estimation. For the latter purpose, a more accurate and less misleading name for the BKG detector could for example be the flare class monitor (FCM).

    \subsection{Future space mission application}

    As demonstrated in this paper, combining the STIX CdTe detectors with a setup of openings as can be found in the BKG detector, is an easy way to monitor the thermal X-ray flux from the Sun. From the perspective of space weather missions, there are several applications for including the Caliste STIX detectors as part of a space weather mission.

    This includes using one or two BKG detectors to measure the thermal X-ray flux. In this case, the name FCM as proposed above would be the more appropriate description. In a simple space weather mission, the detectors would be most likely located at 1 AU. The goal would be to establish a detector setup which covers a range of GOES classes as wide as possible. For the lower end of GOES classes we consider the following calculations: As discussed in this paper, the small apertures only measure sufficient photon counts for large flares. The boundary for the observable GOES classes at 1 AU corresponds to roughly a C7 flare. By decreasing the STIX counts by a factor of ten, we get C1/C2 flares for the middle apertures and B5 flares for the large apertures. This means that at 1 AU, the BKG detector with its current design is sensitive to B5 class flares and above. For the upper limit of GOES classes we can consider the following: The largest flare observed by STIX to date is an estimated X16.5 flare. Using this flare as a conservative limit for the live-time of the middle apertures, the small aperture can provide measurements 10 times more intense in STIX flux. Using the power-law correlation from Eq. \ref{Eq: Result STIX vs GOES}, the enhancement factor for the GOES class becomes 5.2. Hence, an X85 class flare at 1 AU can be accurately measured, or an X15 at 0.3AU (i.e. STIX around perihelion). We marked a potential X85 flare with a green cross in Fig. \ref{Fig: GOES vs STIX}. This estimate is strongly simplified but nevertheless gives a conservative limit. An even larger flare would be more difficult to analyze due to the strongly decreased live-time of the detector.

    In conclusion, the BKG detector on STIX in its current form would be able to measure flares from roughly B5 upwards at 1 AU. Our specific proposition for a new space mission would be to include a similar setup with two BKG detectors. A total of 16 pixels could allow for duplicates of six different sized apertures spaced out in the range 1 mm$^2$ to 0.01 mm$^2$, and four dark pixels to measure the background flux of the instrument. This would provide a better coverage of the smaller flares while preserving the large range of observable GOES-class flares. A general advantage of the BKG detector approach is that there is no need of an attenuator and therefor no moving part in the setup. Instead each pixel can be individually read out or switched off depending on the count rate level.

    In the context of a space weather mission, a STIX Caliste detector with the exact same design but a Silicon sensor instead of the CdTe sensor could be an interesting option. In contrast to the CdTe sensors used in STIX, Silicon sensors can measure down to 1 keV. During a space weather mission this would have the advantage that even lower temperatures in flares are covered. Further, as discussed in the Section \ref{Intro}, the overlap between STIX and GOES could be increased facilitating the comparison between the two instruments.

    \subsection{Conclusion}

    We established and explained the calibration of the BKG detector apertures on STIX onboard Solar Orbiter. The BKG detector monitors the low energy X-ray flux also for the largest flares with inserted attenuator. We used this to describe a relationship between the GOES peak flux and the STIX BKG peak flux. This allows us to determine a flare's estimated GOES class more accurately, especially for the largest flares seen by STIX. Using the BKG detector of STIX as a proxy for the GOES class, our estimate for the largest flare observed by STIX so far is an X16.5 $\pm$ 1.8 on the 20th of Mai 2024.

\begin{acknowledgements}
      \em{We want} to thank the anonymous referee for the comments which improved the overall understand of the paper significantly. We want to thank Marina Battaglia for her valuable input. Solar Orbiter is a space mission of international collaboration between ESA and NASA, operated by ESA. The STIX instrument is an international collaboration between Switzerland, Poland, France, Czech Republic, Germany, Austria, Ireland, and Italy. MZS and SK are supported by the Swiss National Science Foundation Grant 200021L\_189180. AV acknowledges the support of the “Accordo ASI/INAF Solar Orbiter: Supporto scientifico per la realizzazione degli strumenti Metis, SWA/DPU e STIX nelle Fasi D-E”. 
\end{acknowledgements}

% for the bibliography, at the end
\bibliographystyle{aa} % style aa.bst
\bibliography{BKG_Paper} % your references Yourfile.bib

\end{document}